# Enhancing IoMT Security with Explainable Machine Learning: A Case Study on the CICIOMT2024 Dataset

Mohammed Yacoubi[1][0009-0008-2250-5460], Omar Moussaoui[1][0000-0002-0467-7735], Cyril DROCOURT[2][0000-0003-1636-9462]

[1] MATSI Laboratory, ESTO, Mohammed First University, Oujda, Morocco
[2] MIS, UPJV, Amiens, France
```
{mohammed.yacoubi, o.moussaoui}@ump.ac.ma
        cyril.drocourt@u-picardie.fr
```

**Abstract.** Explainable Artificial Intelligence (XAI) enhances the transparency and interpretability of AI models, addressing their inherent opacity. In cybersecurity, particularly within the Internet of Medical Things (IoMT), the black-box nature of AI-driven threat detection poses a significant challenge. Cybersecurity professionals must not only detect attacks but also understand the reasoning behind AI decisions to ensure trust and accountability. The rapid increase in cyber-attacks targeting connected medical devices threatens patient safety and data privacy, necessitating advanced AI-driven solutions. This study compares two ensemble learning techniques, bagging and boosting, for cyber-attack classification in IoMT environments. We selected Random Forest for bagging and CatBoost for boosting. Random Forest helps reduce variance, while CatBoost improves bias by combining weak classifiers into a strong ensemble model, making them effective for detecting sophisticated attacks. However, their complexity often reduces transparency, making it difficult for cybersecurity professionals to interpret and trust their decisions. To address this issue, we apply XAI models to generate local and global explanations, providing insights into AI decision-making. Using techniques like SHAP (Shapley Additive Explanations) and LIME (Local Interpretable Model-agnostic Explanations), we highlight feature importance to help stakeholders understand the key factors driving cyber threat detection.

## 1   Introduction

The Internet of Medical Things (IoMT) is revolutionizing healthcare by enabling continuous monitoring [1] and real-time data access [2]. However, this growing connectivity exposes medical devices to significant cybersecurity threats [3], endangering the confidentiality, integrity, and availability (CIA) of sensitive data and potentially compromising patient safety [4]. As a result, implementing robust attack detection solutions is essential. While ensemble learning techniques like Random Forest [5] and CatBoost have proven effective in cyberattack classification [6]. their complexity makes it challenging to interpret their decisions, which limits their adoption in critical environments such as IoMT [7]. In this paper, we propose a comparative study of Random Forest and CatBoost for cyberattack detection in IoMT environments, integrating XAI techniques such as SHAP and LIME to enhance model interpretability.

The rest of the paper is organized as follows. In the second section, we present backgrounds and explore the related works. Before concluding, the third section discusses the obtained results.

## 2    Backgrounds and Related Works

**IoMT**, a healthcare-focused subset of IoT, connects medical devices to improve patient monitoring, diagnosis, and treatment. However, their resource-constrained nature exposes these systems to threats like DoS, DDoS, Mirai botnet, scanning, Recon, spoofing, and MitM attacks [8]. One of the major challenges in the medical field is intrusion detection to prevent any disruption of patient's medical devices and to protect the healthcare ecosystem from vulnerabilities that could impact various services.

**Machine Learning** (ML) is a collection of data analysis methods that allow machines to learn from experience similarly to how people and animals do [9]. ML primarily aims at performing classification and regression by leveraging known features learned from training data [10]. For this purpose, we selected two ML algorithms: Random Forest (RF) [11] and CatBoost [12].
- Random Forest: An ensemble algorithm combining multiple decision trees, each trained on random subsets of data and features. This diversity reduces overfitting and stabilizes predictions through voting or averaging of results.
- CatBoost: A gradient boosting method optimized for categorical data, automatically handling non-numeric variables. It uses ordered boosting to minimize errors while speeding up training, making it ideal for precise and fast predictions on complex datasets.

**XAI** enhances trust in AI/ML systems by making their outcomes interpretable. It provides tailored explanations for users, operators, and developers to address adoption, governance, and development challenges [13]. This explainable aspect is crucial for fostering the trust necessary for AI to gain widespread market acceptance and deliver its potential benefits. Among related and emerging initiatives, reliable AI and responsible AI are key concepts gaining traction [14]. Explanations of decision models can be classified as either local or global based on the model's scope. Local explainability refers to the ability of a system to clarify for a user the reasoning behind a specific choice or decision. Popular methods for local explainability, such as LIME [15], SHAP [16], fall under this category. These methods are considered essential for achieving model transparency [17]. On the other hand, global explainability involves providing a broader explanation of the learning algorithm as a whole. This includes considering the training data used, the proper application of algorithms, and identifying potential flaws or misapplications[18].

For our analysis, we employed the publicly available CICIoMT2024 dataset from the Canadian Institute for Cybersecurity (CIC) [19]. It includes 46 features across diverse IoT devices (healthcare, home automation, network infrastructure, and auxiliary hardware) and covers 18 distinct cyberattacks.

XAI has become increasingly critical in healthcare [20] and cybersecurity[21]. particularly for intrusion detection in IoMT systems. Frameworks like MetaXAI [22] leverage AI and VR to provide 3D explanations via SHAP, LIME, and ELI5, achieving a 94% success rate. This study compares RF and CatBoost models, analyzing their feature importance extraction using SHAP and LIME and evaluating their efficiency in generating global and local explanations.

## 3      Results and Discussion

This study aims to enhance cyberattack detection in IoMT systems using ensemble learning. We compare RF and CatBoost based on evaluation metrics and training time.

Experiments were conducted with Python 3.11.11 on an Intel Core i5-8250U CPU. CatBoost outperformed RF in handling large datasets, benefiting from categorical data optimization and parallelized gradient boosting.

Both models were trained on a 1M row dataset, with 70% used for training and 30% for testing. Table 1 summarizes performance results, while Table 2 details model parameters. Finally, SHAP and LIME were used to analyze feature impact on predictions.

Table 1. Evaluation results of the ensemble models.

| Model | Accuracy | Precision | Recall | F1-score | ROC AUC | Time Training |
|---|---|---|---|---|---|---|
| RF | 99.92% | 99.97% | 99.95% | 99.96% | 100.00% | 140 s |
| CatBoost | 99.90% | 99.97% | 99.93% | 99.95% | 100.00% | 23 s |

Table 2. models details and hyperparameters.

| Model | Hyperparameters |
|---|---|
| RF | n_estimators=100, criterion="gini", min_saples_split=2, min_saples_leaf=1, min_weight_fraction_leaf=0.0, max_fetures="sqrt", min_impurity_decrease=0.0, bootstrap=True, oob_score=False, warm_start=False, ccp_alpha=0.0, random_state=42. |
| CatBoost | iterations=100, depth=6, learning_rate=0.1, verbose=0. |

Here, we present extended results of XAI techniques applied to our AI models. Figure 1 illustrates feature importance using SHAP-based global beeswarm plots for the CICIOMT2024 dataset, showing the correlation between feature values and model predictions. For Random Forest, the most important features are positioned on the left, with "IAT" being the highest, followed by "rst_count", while the least important feature, "Srate", appears at the bottom. In contrast, for CatBoost, the most important features are "Rate" and srate, whereas "psh_flag_number" remains the least important among the top 10 features. Figures 2 and 3 demonstrate local interpretations using the LIME framework, where the RF and CatBoost models, respectively, accurately predict attack traffic instances from the CICIOMT2024 dataset.

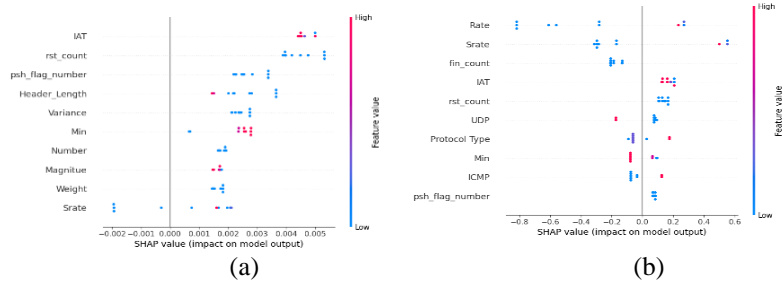

**Fig. 1.** Feature importance using SHAP-based global beeswarm plots for the CICIOMT2024 dataset. The global beeswarm plot illustrates the correlation between feature values and the model's predictions. (a) RF, (b) CatBoost.

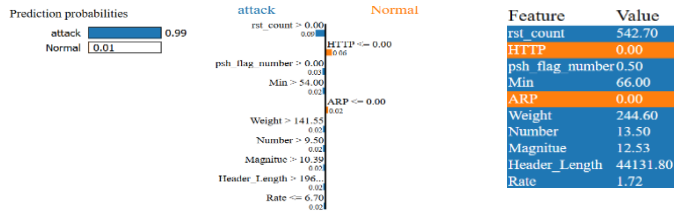

**Fig. 2.** A demonstration of a local interpretation, where the RF model accurately predicts an attack traffic instance from the CICIOMT2024 dataset, is provided using the LIME framework.

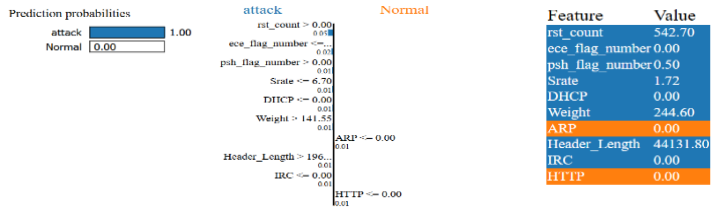

**Fig. 3.** A demonstration of a local interpretation, where the CatBoost model accurately predicts an attack traffic instance from the CICIOMT2024 dataset, is provided using the LIME framework.

**Table 3.** Runtime and Sample Size Comparison of LIME and SHAP for RF and CatBoost Models

| Model | LIME | | SHAP | |
|---|---|---|---|---|
| | Nombre sample | Execution Time (s) | Nombre sample | Execution Time (s) |
| RF | 1 | 13 | 1000 | 77 |
| CaBoost | 1 | 13 | 1000 | 3.5 |

A key aspect of XAI is fast data explanation, essential for real-time applications and dependent on ML model selection. Table 3 compares SHAP and LIME for RF and CatBoost, showing that SHAP is highly effective for global explanations with Boosting models, achieving 3.5s for 1,000 instances. Meanwhile, LIME plays a critical role in local explanations for identifying predicted classes.

## 4    Conclusion

This study highlighted the critical impact of XAI techniques, particularly SHAP and LIME, in enhancing the interpretability and reliability of ensemble models for cyberattack detection in IoMT systems. The results underscored CatBoost's computational superiority, especially with SHAP, which reduced processing time by a factor of 22-fold compared to Random Forest. Meanwhile, LIME achieved ultra-fast local interpretations 1 sample in 13s, aligning with real-time constraints.

In future work, we will explore supervised wrapper and embedded methods for feature selection, evaluating their speed and compatibility with XAI. The objective is to investigate whether XAI can emerge as an innovative feature selection methodology, optimizing both performance and interpretability in cybersecurity applications.